# Transverse Velocities of Galaxies From Microlens Parallaxes

**Andrew Gould**

Dept of Astronomy, Ohio State University, Columbus, OH 43210

e-mail gould@payne.mps.ohio-state.edu

**Abstract**

The transverse velocity of a spiral galaxy can be measured to an accuracy $\sim 60\,\mathrm{km\,s^{-1}}$ by making parallax observations of quasars being microlensed by stars in the disk of the galaxy. To make the measurement, a quasar must be located behind the disk of the galaxy between about 1 and 2 scale lengths from the center. The quasar must then be monitored for microlensing events and the events followed simultaneously from the Earth and a satellite in solar orbit, preferably at $\sim 30\,\mathrm{AU}$. A systematic search in a volume within $21,000\,\mathrm{km\,s^{-1}}$ could locate quasars with $B < 23$ behind a total of $\sim 1900$ galaxies. The rate of lensing events (and hence galaxy velocity measurements) would be $\sim 3\,\mathrm{yr^{-1}}$. The events would have typical characteristic times $\omega^{-1} \sim 3\,\mathrm{yr}$. Under the assumption that the mass spectrum of lensing objects is the same in other spiral galaxies as in our own, the observations could be used to measure the Hubble parameter to an accuracy of 5%.

Subject Headings: astrometry – gravitational lensing

## 1. Introduction

Measurement of the transverse velocities of galaxies could provide important information about the large scale structure of the universe. For example, Dekel, Bertschinger, & Faber (1990) have shown that one may reconstruct the local galaxy velocity field (including the transverse components) from measurements of peculiar radial velocities, provided that one assumes that the velocity field is irrotational. By measuring the transverse velocities one could test this assumption. Several workers have attempted to measure the bulk flow of the nearby universe out to various redshift limits from 3000 to 15,000 $\rm km\,s^{-1}$ (Courteau et al. 1993; Mould et al. 1993; Lauer & Postman 1994). In particular, they have sought but so far failed to find a volume sufficiently large that it is stationary with respect to the cosmological reference frame defined by the microwave background radiation. The measurement of transverse velocities would yield an independent determination of these bulk flow velocities.

In addition, all peculiar radial velocity measurements are based on the comparison of a redshift with a distance as determined from a distance indicator. This approach has two intrinsic drawbacks. First, since the distance errors are proportional to the distance, the fractional errors in the peculiar velocity are also proportional to the distance. Thus, even if the distance errors are only $\sim 6\%$ as claimed for the method of surface brightness fluctuations (Tonry 1991), the errors rise to order unity for $600\,\rm km\,s^{-1}$ peculiar velocities measured at redshifts of $10,000\,\rm km\,s^{-1}$. Other methods, such as Tully-Fisher for spirals and $D_n$-$\sigma$ for ellipticals have substantially larger errors and hence smaller effective ranges. One would like to have a method of measuring velocities for which the errors are independent of distance. Second, all secondary distance-indicator measurements are based on correlations between various observed properties of galaxies. The physical bases of these correlations are incompletely understood. One must therefore worry that subtle selection effects might introduce systematic errors in the distance determinations. Even small systematic errors in the fractional distance measurement can be serious at large distances where the peculiar velocity is itself a small fraction of the redshift. One would like to have a method of measuring velocities that is based directly on known physical principles.

Here I present a method of measuring the transverse velocities of galaxies from the parallax of microlensing events generated by stars within the galaxy. The method is based on simple and well-established physical principles. For spiral galaxies, the accuracy of the measurement is limited only by the dispersion of stars in the disk, typically $\sim 40\,\mathrm{km\,s^{-1}}$ in each transverse direction. In particular, the limiting errors are independent of distance. The measurements can be carried out using existing technologies. They do however require a considerable investment of resources including a $\sim 1$m telescope in solar orbit, preferably at $\sim 30\,\mathrm{AU}$, as well as an intensive search for faint quasars behind thousands of spiral galaxies.

Grieger, Kayser, & Refsdal (1986) were the first to suggest that the transverse velocities of distant galaxies could be estimated using gravitational microlensing of distant quasars. They pointed out that if a quasar is highly magnified by a lensing galaxy, then the stars in the galaxy produce a network of microlensing caustics. If the line of sight from the Earth to the quasar should pass over one of these caustics, the magnification would reach very high levels. By comparing the times of maximum magnification as seen from the Earth and a satellite, one could measure the "reduced" transverse speed up to an unknown geometric factor. The reduced velocity $\tilde{\mathbf{v}}$ is related to the true transverse velocity $\mathbf{v}$ of the lens relative to the observer-source line of sight by

$$\tilde{\mathbf{v}} = (1+z)\frac{D_{\mathrm{OS}}}{D_{\mathrm{LS}}}\mathbf{v}, \qquad (1.1)$$

where $D_{\mathrm{OS}}$, $D_{\mathrm{LS}}$, and $D_{\mathrm{OL}}$ are the angular diameter distances between the observer, source, and lens.

I have previously shown that the reduced transverse velocity of Massive Compact Objects (MACHOs) can be measured by comparing the microlensing light curves as seen from the Earth and a satellite (Gould 1992, 1994a, 1994b). Note that for MACHOs both components of $\tilde{\mathbf{v}}$ can be measured without ambiguity, whereas for the highly magnified quasar only the reduced speed can be measured and then only up to an unknown correction. This important difference is rooted in the topologies of the caustic structure of each type of lens. MACHOs are isolated point masses and so induce a circularly symmetric magnification field about an isolated point caustic. The difference in light curves therefore gives two dimensional information about the motion of the lens relative to the Earth-satellite separation vector (see Figs. 1 and 2 in Gould 1994a). By contrast, the highly magnified quasar field is laced with *linear* caustics. Hence, information about relative motions is inherently one dimensional (see Fig. 1 in Grieger et al. 1985).

The method I propose here is simply to search for non-highly magnified quasars behind the disks of spiral galaxies. The stars in such galaxies give rise to isolated, circularly symmetric magnification fields. The method developed for measuring MACHO reduced velocities therefore applies directly to such lensing events. In the case of distant quasars lensed by stars in relatively nearby galaxies, $(1+z)D_{\rm OS}/D_{\rm LS}$ is close to unity and can in any event be accurately measured. Hence the transverse velocity can be inferred from the measurement of the reduced velocity. This situation is much superior to the MACHO case where the ratio $D_{\rm OS}/D_{\rm LS}$ is generally unknown.

The reader who is interested in obtaining a clear picture of the procedure for determining reduced velocities from parallax measurements is urged to study Gould (1994a, 1994b).

I propose an experiment that would monitor quasars brighter than $B \sim 23$ behind $\sim 1900$ galaxies over $\sim 3\pi$ steradians out to a redshift $\sim 21,000\,{\rm km\,s^{-1}}$. I estimate that one could measure $\sim 3$ transverse velocities per year i.e., $\sim 100$ over three decades. Such a large sample of high precision measurements would both serve as an independent check on models derived from radial measurements and greatly enhance the accuracy of the models. For example, assuming that the galaxies themselves have random motions $\lesssim 200\,{\rm km\,s^{-1}}$ in each direction, such a sample would measure the bulk flow of the sampled volume with a precision $\lesssim 25\,\mathrm{km\,s^{-1}}$. The bulk flow of each of 10 subregions could be measured to $\lesssim 80\,\mathrm{km\,s^{-1}}$. These determinations would be free from the "rocket effect" (Kaiser 1987) that afflicts studies based on radial peculiar velocities.

In addition, the experiment would have several other important applications. It could be used to measure the Hubble parameter, to determine the ratio of stellar mass to light of spiral disks, and to study the structure of quasars on scales of 100–1000 AU.

## 2. Lensing Events

Suppose that a distant quasar is lensed by a star in a relatively nearby galaxy at a distance $D = D_{\mathrm{OL}}$. I assume that the optical depth $\tau$ for lensing by stars along the line of sight through the galaxy is low

$$\tau = \frac{4\pi G \Sigma D_{\mathrm{OL}} D_{\mathrm{LS}}}{D_{\mathrm{OS}} c^2} \to \frac{4\pi G \Sigma D}{c^2} \ll 1, \tag{2.1}$$

where $\Sigma$ is the column density of stars and where I have assumed $D_{\mathrm{LS}}/D_{\mathrm{OS}} \simeq 1$. Equation (2.1) can be rewritten as $\pi \Sigma / \Sigma_{\mathrm{crit}} \ll 1$ where $\Sigma_{\mathrm{crit}}$ is the critical density. Hence $\tau \ll 1$ is equivalent to the surface density being well below the critical level required for macrolensing. In turn, this usually implies that the shear is low. In any event, I will assume that this is the case. Hence, the star can be treated as an isolated microlens and the formalism that I have previously developed (Gould 1994a) can be applied directly. The Einstein ring radius, $r_e$ is given by

$$r_e = \left(\frac{4GMD}{c^2}\right)^{1/2} = 280\,\mathrm{AU}\left(\frac{M}{0.1 M_\odot}\right)^{1/2}\left(\frac{D}{D_{\mathrm{Coma}}}\right)^{1/2}, \tag{2.2}$$

where $M$ is the stellar mass and $D_{\mathrm{Coma}}$ is the distance to the Coma cluster which I take to be 100 Mpc (i.e. $H_0 = 70\,\mathrm{km\,s^{-1}\,Mpc^{-1}}$). The Einstein ring projected onto the source plane has size $(D_{\mathrm{OS}}/D_{\mathrm{OL}}) r_e \sim 10^4\,\mathrm{AU}\,(M/0.1 M_\odot)^{1/2}(D/D_{\mathrm{Coma}})^{-1/2}$,

where I have assumed that the quasar is at a cosmological distance. This size is much larger than the size of the quasar continuum source. Therefore, quasars that are much brighter than their hosts can be treated as point sources. The lensing event as viewed from the Earth can be described by a three parameter light curve

$$A[x(t)] = \frac{x^2 + 2}{x(x^2 + 4)^{1/2}}; \qquad x(t) = \sqrt{\omega^2(t - t_0)^2 + \beta^2}, \qquad (2.3)$$

where $A(x)$ is the magnification, $x$ is the projected separation of the lens and source in units of the Einstein radius, $\omega^{-1}$ is the characteristic event time, $t_0$ is the time of maximum magnification, and $\beta$ is the dimensionless impact parameter. The light curve as seen from a satellite at position $\mathbf{r}$ relative to the Earth has a similar form but with different parameters $\beta'$, $t_0'$ and $\omega' \simeq \omega$. One can form a vector $\Delta \mathbf{x} \equiv (\omega \Delta t, \Delta \beta)$ where $|\Delta \beta| = |\beta' \pm \beta|$ and $\Delta t = t_0' - t_0$. Ignoring for the moment the four-fold ambiguity in $\Delta \beta$ (see § 5.3), one can determine the transverse velocity by measuring the parameters of each light curve and using the relation $\mathbf{v} = \omega r \Delta \mathbf{x}/(\Delta x)^2$. Note that $\Delta x = r/r_e$ should be in the range $0.1 \lesssim \Delta x \lesssim 1$ in order to make a reasonably accurate ($\lesssim 10\%$) velocity measurement. Hence, to measure the transverse velocities of galaxies out to three times the distance of Coma, the satellite orbit should be at least $\sim 30\,\mathrm{AU}$. The characteristic time of a typical event would be

$$\omega^{-1} = \frac{r_e}{v} = 2.2\,\mathrm{yr} \left(\frac{M}{0.1 M_\odot}\right)^{1/2} \left(\frac{D}{D_{\mathrm{Coma}}}\right)^{1/2} \left(\frac{v}{600\,\mathrm{km\,s^{-1}}}\right)^{-1}. \qquad (2.4)$$

# 3. Proposed Experiment

In order to measure the transverse velocity of a spiral galaxy, one must first find a quasar behind the galaxy's disk. The number density of quasars brighter than $B_0 = 22$ is $n_{22} \sim 100 \deg^{-2}$ (Hartwick & Schade 1990). The number brighter than $B_0 = 23$ is $n_{23} \sim 150 \deg^{-2}$ (P. Osmer 1994, private communication). To estimate the number of galaxies with quasars behind them, I will assume that all galaxies are face on and that quasars with $B_0 < 23$ can be detected between 1 and 2 disk scale lengths. I take M31 to be a 'typical' $L^*$ spiral, and will assume that there are $1.2 \times 10^{-2} h^3 \text{Mpc}^{-3} \sim 4 \times 10^{-3} \text{Mpc}^{-3}$ such galaxies (Binney & Tremaine 1987) with scale length $h = 6.4 \text{ kpc}$ and total blue luminosity of $L = 2.4 \times 10^{10} \, L_\odot$ (van der Kruit 1989). I take the stellar mass to blue light ratio to be $4 \, M_\odot/L_\odot$, implying a surface stellar density at radius $\ell$ of

$$\Sigma = 375 \exp(-\ell/h) \, M_\odot \, \text{pc}^{-2}. \tag{3.1}$$

The adopted assumptions are obviously rather crude. I justify this approach below.

I then find that of galaxies at a distance $D$ a fraction $f = 3\pi(h/D)^2 n_{23} = 0.019(D/D_{\text{Coma}})^{-2}$ have quasars behind them between 1 and 2 scale lengths. For those galaxies with quasars at radius $\ell$, $\tau = 0.023 \exp(-\ell/h)(D/D_{\text{Coma}})$. Hence the mean optical depth over the disk is $\tau = 0.005 \, (D/D_{\text{Coma}})$. The rate of lensing events by an average galaxy with a quasar behind it is then $\Gamma = (2/\pi)\tau\omega$ or

$$\Gamma = 1.4 \times 10^{-3} \, \text{yr}^{-1} \left(\frac{M}{0.1 M_\odot}\right)^{-1/2} \left(\frac{D}{D_{\text{Coma}}}\right)^{1/2} \left(\frac{v}{600 \, \text{km s}^{-1}}\right). \tag{3.2}$$

Consider now a volume limited sample restricted to the $3\pi$ steradians ($|b| \gtrsim 20°$) within the galactic polar caps and inside a distance $D_{\max}$. The total number of $L^*$ galaxies in the sample is $N = 12500(D/D_{\text{Coma}})^3$. Of these, a total of

$$N_Q = 700 \frac{D_{\max}}{D_{\text{Coma}}}, \tag{3.3}$$

will have quasars behind them. The rate of lensing events will then be

$$\Gamma_{\rm tot} = 0.68\,{\rm yr}^{-1} \left(\frac{M}{0.1 M_\odot}\right)^{-1/2} \left(\frac{D_{\rm max}}{D_{\rm Coma}}\right)^{3/2} \left(\frac{v}{600\,{\rm km\,s}^{-1}}\right). \qquad (3.4)$$

In deriving these results I assumed that all galaxies are face on. If the galaxies are at all inclinations but all the quasars behind the galaxies are located, the total number of quasars will be reduced but the optical depth for each will be proportionately increased. Hence, equation (3.4) remains valid. Of course, if a galaxy is highly inclined, the extinction through the inner disk will be increased and some of the background quasars will be lost. On the other hand, the disk will then have a relatively high optical depth even at several scale lengths so that it would be profitable to search a greater part of the disk for quasars. I also assumed that $B_0 = 23$ quasars could be found even at 1 scale length where the extinction might be $A_B \sim 1$, so that $B \sim 24$. As I will discuss in the next section, finding and monitoring quasars that faint may prove difficult. On the other hand, the number of quasars $B_0 < 22$ within the inner scale length will not be insignificant and these (with their higher optical depth) will compensate for the lost faint quasars further out. At low latitudes some of the fainter quasars will be lost due to foreground extinction, but this results in only a $\sim 10\%$ reduction in the sample averaged over all angles. Because the quasar luminosity function is not rising rapidly at these apparent magnitudes, the estimates that I have made are not very sensitive to the exact cutoff assumed.

From equations (3.3) and (3.4) and taking account of the 10% reduction due to foreground extinction, I find that a survey out to $D_{\rm max} = 3 D_{\rm Coma}$ ($\sim 21{,}000\,{\rm km\,s}^{-1}$) would monitor $N_Q \sim 1900$ quasars and measure the transverse velocities of $\sim 3$ galaxies per year. After 30 years, $\sim 100$ velocities would be measured, implying a mean sampling scale of $\sim 6600\,{\rm km\,s}^{-1}$ over the whole volume. The measurement errors would be $\mathcal{O}(10\%)$ in each direction (see § 5.3). For typical expected transverse velocities $\sim 400\,{\rm km\,s}^{-1}$ in each direction, this error is comparable to the intrinsic error arising from the rms difference between the lensing star's velocity and the mean velocity of the spiral disk at its location. Hence the overall errors would be $\sim 60\,\rm km\,s^{-1}$. This is less than the expected random motions of galaxies relative to the mean peculiar velocity flow.

## 4. Scientific Byproducts

Here I briefly describe the scientific questions which could be addressed with such a survey other than the results pertaining to large-scale structure described in §1.

### 4.1. Measurement of the Hubble Parameter

Each transverse-velocity measurement will also yield $r_e$, the radius of the Einstein ring induced by the lensing star. From equation (2.2) and Hubble relation one finds $M = r_e^2 c/4GH_0 z$ where $H_0$ is the Hubble parameter and $z$ is the galaxy redshift. Hence the mass spectrum of the lensing stars in an ensemble of galaxies can be measured up to an overall factor of $H_0$. The errors in the individual mass measurements would have errors ($\sim 20\%$) that are small compared to the half-width of the mass spectrum (which I take to be $\sim 50\%$). The mass spectrum of lensing stars in the Galactic disk can be determined by a similar experiment (in many ways a small prototype for the one proposed here) that would measure parallaxes of MACHO events seen in ongoing lensing experiments (Gould 1994a, 1994b, Han & Gould 1994). Hence, if one assumed that the Galactic spectrum was the same as the composite extragalactic spectrum, the Hubble parameter could be measured to $\sim 5\%$ from 100 events. Of course, the mass spectrum may vary according to galaxy type or position in the galaxy. However, the variation of the spectrum as a function of galactocentric radius in the Milky Way can be measured from MACHO parallaxes (Han & Gould 1994), and a similar test can be performed by dividing the extragalactic sample into subsamples according to the radial position of the quasar line of sight. Similarly, the extragalactic sample can be divided up into subsamples of different galaxy types to test for different mass functions.

## 4.2. Ratio of Stellar Mass to Light

The experiment allows one to measure the mean value of the surface mass density along all the lines of sight to program quasars by combining the observed mean optical depth with the galaxy distance distribution. This can be compared directly with the mean surface brightness to get a ratio of stellar mass to observed light. One could also find the correction for self absorption by studying the optical depth versus surface brightness as a function of galaxy inclination.

## 4.3. Central Regions Of Quasars

As I discussed in § 2, the Einstein ring projected onto the source plane will have a radius of order $\sim 10^4$ AU, so that the form of the lensing event will generally not be affected by the much smaller structure of the quasar source region. However, a fraction of events $\beta_*$ will have impact parameters $\beta < \beta_*$ and hence will come within $10^4 \beta_*$ AU of the central source. Thus over 30 years, 10 events will come within 1000 AU and 1 within 100 AU. These regions can then be probed with ground-based spectra and or analysis of the deviation of the light curves from the expected microlensing form.

# 5. Practical Requirements

Here I give a brief description of some of the potential obstacles to carrying out the experiment and how these obstacles might be overcome.

## 5.1. Quasar Identification

The surface brightnesses of spiral disks are fainter than the sky, so it is no more difficult to locate a quasar behind a galactic disk than it is to find an object of the same apparent magnitude in the field. However, there are potential problems posed by contamination.

Possible contaminants are foreground stars, faint galaxies, and stars, globular clusters, and HII regions in the spiral galaxy. Quasars with $z < 2.2$ generally have a substantial UV excess. If this is the principal method by which the quasars are selected, then virtually all globular clusters, HII regions, and stars in the galaxy will be excluded. All Galactic foreground stars except young disk white dwarfs and distant horizontal branch stars will likewise fail the test. Horizontal branch stars at $B \sim 23$ are so distant ($\sim 250 \, \mathrm{kpc}$) that they are extremely rare. Extremely hot white dwarfs are also rare and moreover, can be found by their parallax offset as seen from the Earth and the satellite. Most background galaxies can be eliminated by the color criterion and most of the rest can be resolved in good ground-based seeing. Experience with the *Hubble Space Telescope* (e.g. Bahcall et al. 1994) shows that a few hours of observations on a 1m satellite telescope could easily resolve the remaining ambiguities. These observations could be undertaken as the satellite traveled toward its Neptune like orbit.

In any event, contamination at the level of a few tens of percent is not critical. Ground based spectra could be taken of the brighter sources and the fainter ones could simply be left in the sample and monitored over the first few years for variation.

### 5.2. Event Identification

Some indication that an event may be going on will come from the light curve (which to first order will be the same as seen from the Earth and the satellite). However, since quasars are intrinsically variable, this signature would not be definitive. Absolute proof of lensing can come only by demonstrating that the ground-based and space-based light curves are different. Regardless of any intrinsic variation, the fractional difference between the two curves will be $\Delta A = 8(r/r_e)\cos\theta/[x(x^2+2)(x^2+4)]$ where $\theta$ is the angle between the projected Earth-satellite and source-lens vectors. Since $r/r_e \sim \mathcal{O}(10\%)$, this effect is $\Delta A \sim \mathcal{O}(2\%)$ at $x \sim 1.3$. Since events have characteristic times $\omega^{-1} \sim 3\,\mathrm{yr}$, the effect would be noticed provided that $\sim 3\%$ photometry were carried out from the satellite once every other month. Once it was noticed, a definitive set of three 1% measurement could be made to confirm the lensing character of the event.

### 5.3. Measurement of Velocity

An important difference between measuring the reduced velocities of a Galactic MACHO against a background star and of an extragalactic lens against a background quasar is that the latter source is variable. Hence, the overall light curves as seen from both the Earth and the satellite deviate from standard microlensing light curves and the accuracy with which the basic parameters $\omega$, $\beta$, and $t_0$ can be measured from these curves may be significantly compromised. However, the ratio of the fluxes seen by the Earth and satellite is independent of the quasar's variability. Moreover, it is possible in principle to reconstruct $\omega$, $\Delta t$, $\beta$ and $\beta'$ from the ratio of light curves. Unfortunately, for a given level of photometry, the errors are then much greater than if one could measure the unadulterated microlensing curves directly.

To investigate this problem quantitatively, I perform the following calculations. I assume that events are typically recognized at separations of 1.3 Einstein radii. Before this time, I assume that the satellite makes 3% photometry measurements every other month and afterward it makes 1% measurements every week (as required to break the degeneracy among the 4 possible solutions – see below). The ground-based measurements are assumed to have substantially smaller errors. I consider the relatively difficult case of a galaxy at the edge of the volume, with $r_e = 350\,\mathrm{AU}$ and $\omega^{-1} = 4\,\mathrm{yr}$. I then estimate the errors in measuring $\omega$, $\omega\Delta t$, and $\Delta\beta_\pm$ under two sets of assumptions: first, that some information about $\omega$ and $\beta$ are available from the overall light curve and second, that the parameters are estimated only from the ratio of the two curves. I find that for $\beta \lesssim 0.5$ and for many orientations of $\mathbf{v}$ relative to $\mathbf{r}$, it is possible just from the light-curve ratio to measure $\Delta\mathbf{x}$ to $\lesssim 10\%$ and $\omega$ to $\lesssim 20\%$. The latter value then dominates the errors. However, the situation deteriorates sharply if $\mathbf{v} \cdot \mathbf{r} \sim 0$. In addition, for $\beta \sim 1$ the errors rise to order unity at all orientations. On the other hand, if the distortions of the overall light curve due to quasar variability are not severe so that the time scale $\omega^{-1}$ can be estimated to $\sim 10\%$ then the typical errors even at $\beta \sim 1$ are only $\sim 15\%$.

At present therefore, the largest single uncertainty in outlining the proposed experiment is our poor knowledge of the statistics of quasar variability. After the MACHO project (Alcock et al. 1993, K. Cooke 1994, private communication) identifies the quasars in their Large Magellanic Cloud fields, it will be possible to make a systematic study of quasar variation and to estimate the effects on the galaxy velocity measurement. Here I will give only a brief overview of the velocity determination. I will assume (as I believe to be the case), that quasar variability does not qualitatively compromise the measurement of the overall light-curve parameters.

The accuracy of the velocity measurement then depends on the accuracy of $\Delta \mathbf{x} = (\omega \Delta t, \Delta \beta)$. The difficulty of determining $\Delta \mathbf{x}$ is greatly reduced by the fact that typically $r << r_e$. This means that for $D \gtrsim D_{\text{Coma}}$ both light curve trajectories usually pass on the same side of the lens. That is, $|\Delta \beta| = \Delta \beta_-$ (see Fig. 1 from Gould 1994a). For example, if $\beta \sim \beta' \sim 0.5$ and if one assumed that the curves passed on opposite sides of the lens, then from equation (2.2), the mass would be $M \lesssim 1 \times 10^{-3} \, M_\odot$. If the mass (and Einstein ring) were this small, it would be easy to tell as I show below. However, since such events are rare, I do not consider them in detail here. Another class of rare events is when $\beta, \beta' \ll 1$. In this case it may be that $|\Delta \beta| = \Delta \beta_+$.

I focus on the typical case where $|\Delta \beta| = \Delta \beta_-$. The main problem is distinguishing between the $\pm \Delta \beta_-$ solutions. This degeneracy can be broken by the effect of the Earth's orbit. The ratio of light curves will oscillate annually as the Earth-source line of sight moves nearer and further from the projected MACHO position (see Fig. 2 in Gould 1992). The time of the largest positive fluctuation differs by about 6 months between the $+\Delta \beta_-$ and $-\Delta \beta_-$ solutions. The difference between the curves for the two solutions has an amplitude $\sim 2\text{AU}/\beta r_e$, i.e., $\sim 0.4\%$ at the edge of the volume for $\beta = 1$. Hence they could be distinguished at the $3\sigma$ level if 1% photometry were carried out once per weak. This required level of accuracy determines the minimum characteristics of the telescope. At any given time there will be $\sim 30$ events that must be followed. Assuming a typical apparent magnitude of $V \sim 23$, it would take $\sim 60\,\text{hr/week}$ of 1 meter telescope time (including overhead) to get 1% photometry on these objects each week. An additional 60 hr/week would be required to get 3% photometry on the remaining 1900 program objects every other month. Thus the telescope must be close to 1m in diameter.

The motion of the Earth can easily distinguish between the $\Delta\beta_\pm$ solutions in the case mentioned above when $\Delta\beta_- \ll 1$ and $\Delta\beta_+ \sim \mathcal{O}(1)$. The latter solution has an Einstein ring which is smaller by a large factor. Hence the amplitude of the annual oscillation is much larger. Since the experiment is designed to detect the smaller oscillations at the $3\,\sigma$ level, this large effect would clearly stand out.

## 6. Conclusions

An experiment to measure the transverse velocities of galaxies with an accuracy of $\sim 60\,\text{km}\,\text{s}^{-1}$ at a rate of $\sim 3\,\text{yr}$ appears to be generally feasible. A 1m telescope in a Neptune-like orbit would be required. The experiment would yield information on large scale structure that would largely be free of systematic errors and so would provide a powerful test of theories. In addition, it could be used to measure the Hubble parameter to $\sim 5\%$, to measure the stellar mass-to-light ratio of galaxies, and to probe the central regions of quasars. The largest uncertainty in designing the experiment is the effect of quasar variability. A variability study of quasars found in the MACHO observations (Alcock et al. 1993) now underway should allow detailed modeling of this effect.

**Acknowledgements**: I would like to thank K. Freeman, P. Osmer, B. Peterson, and P. Schneider for valuable discussion.

# REFERENCES

Alcock, C., et al. 1993, Nature, 365, 621

Bahcall, J. N., Flynn, C., Gould, A. & Kirhakos, S. 1994, ApJ Letters, in press

Binney J. & Tremaine, S. 1987, Galactic Dynamics, p. 22, (Princeton: Princeton Univ. Press)

Courteau, S., Faber, S. M., Dressler, A., & Willick, J. A. 1993, ApJ, 354, 18

Dekel, A., Bertschinger, E., & Faber, S. M. 1990, ApJ, 364, P349

Gould, A. 1992, ApJ, 392, 442

Gould, A. 1994a, ApJ, 421, L75

Gould, A. 1994b, ApJ Letters, submitted

Grieger, B., Kayser, R., & Refsdal, S. 1986, Nature, 324, 126

Han, C. & Gould, A. 1994, in preparation

Kaiser, N. 1987, MNRAS, 227, 1

Lauer, T. R. & Postman, M. 1994, ApJ, 425, 418

Mould, J. R., Akeson, R. L., Bothun, G. D., Han, M., Huchra, J. P., Roth, J., & Schommer, R. A. 1993, ApJ, 383,467

Tonry, J. 1991, ApJ, 373, L1

van der Kruit, P. C. 1989, The Milky Way As A Galaxy, R. Buser and I. R. King, eds., (Mill Valley: University Science Books)